\documentclass[%
 reprint,
preprintnumbers,
nofootinbib,
 amsmath,amssymb,
 aps,
]{revtex4-2}

\usepackage{lineno}
\usepackage[breaklinks]{hyperref}
\usepackage{amsmath}
\usepackage{url,xurl}            
\usepackage{breakurl} 
\usepackage{cleveref}
\usepackage{booktabs}       
\usepackage{amsfonts}       
\usepackage{nicefrac}       
\usepackage{microtype}      
\usepackage{braket}
\usepackage{macros}
\usepackage{subfigure} 
\usepackage{color}
\usepackage{comment}
\usepackage{float}
\usepackage{bbm}
\usepackage{graphicx}
\usepackage{dcolumn}
\usepackage{bm}
\usepackage{ulem}
\usepackage{slashed}

\begin{document}

\preprint{TTK-23-31}

\title{Moments of parton distribution functions 
of any order from lattice QCD}

\author{Andrea Shindler}
\email{shindler@physik.rwth-aachen.de}
\affiliation{%
Institute for Theoretical Particle Physics and Cosmology, 
 TTK, RWTH Aachen University}%
\affiliation{%
Nuclear Science Division, Lawrence Berkeley National Laboratory, Berkeley, California 94720, USA}%
\affiliation{%
Department of Physics, University of California, Berkeley, California 94720, USA}%

\date{\today}

\begin{abstract}
    We describe a procedure to determine moments of parton distribution 
    functions of any order in lattice quantum chromodynamics (QCD). The procedure is based on the gradient flow 
    for fermion and gauge fields. The flowed matrix 
    elements of twist-2 operators renormalize multiplicatively, 
    and the matching with the 
    physical matrix elements can be obtained using continuum 
    symmetries and the irreducible representations 
    of Euclidean $4$-dimensional rotations. We calculate
    the matching coefficients at one-loop in perturbation theory 
    for moments of any order in the flavor nonsinglet case. 
    We also give specific examples 
    of operators that could be used in lattice QCD computations.
    It turns out that it is possible to choose operators 
    with identical Lorentz indices 
    and still have a multiplicative matching. One can thus
    use twist-2 operators exclusively with temporal indices, 
    thus substantially improving the 
    signal-to-noise ratio in the computation of the hadronic matrix elements.
\end{abstract}

\maketitle

{\it{Introduction.}} 
Quantum chromodynamics (QCD) describes quark-gluon interactions over 
a wide energy range, posing nonperturbative challenges at lower energies. 
Hadron structure studies depend on parton distribution functions (PDFs), 
with PDF determinations relying on precise QCD calculations, 
influencing investigations at the LHC and future projects (EIC, HL-LHC, LHeC).
The quest for new physics hinges on accurate experimental-theoretical comparisons, 
emphasizing PDF precision.

QCD factorization in collider processes, beyond PDFs, 
involves generalized parton distributions (GPDs) and 
transverse-momentum-dependent distributions (TMDs). 
While not directly measurable, these distribution functions 
can be determined by factorizing the process's cross section 
into hard and soft components.

In deep inelastic scattering (DIS), a key role is played by the structure 
functions, $F_k(x,Q^2)~(k=1,2,3)$, which depend on $Q^2=-q^2$, where $q$ is the 4-momentum 
transferred to the nucleon, and $x = Q^2/2(pq)$ represents 
the fraction of nucleon momentum carried by the parton. 
Here, $p$ denotes the nucleon momentum.
For $Q^2$ much larger than the nucleon mass, 
the factorization theorem~\cite{Collins:1989gx} 
relates the structure functions to the convolution of 
the scale-dependent parton distribution functions $f_a(x,\mu^2)$ 
with the Wilson coefficients $C_k^a(Q^2/\mu^2)$ 
(summed over quark flavors and gluons, $a$).
The Wilson coefficients are calculable in perturbation theory, and the 
evolution of the PDFs in $\mu$ is governed by the 
Dokshitzer-Gribov-Lipatov-Altarelli-Parisi (DGLAP) 
equations~\cite{Gribov:1972ri,Lipatov:1974qm,Dokshitzer:1977sg,Altarelli:1977zs}.
The DGLAP equations can be solved once we specify an initial condition at 
a certain scale $\mu=\mu_0$. The evolution of the PDFs is calculable 
in perturbation theory, but the precision of experimental data 
always necessitates higher-order calculations. 
The standard choice for renormalization 
and factorization scheme is the $\MSbar$ scheme.
PDFs can be determined experimentally by analyzing a set of 
hard-scattering processes involving nucleons.
There is a long-standing community effort to determine 
PDFs using well-defined parametrizations for their $x$-dependence
and starting the evolution around $\mu_0^2 = 1 - 4$GeV.

Lattice QCD can offer a theoretical input for the determination of the 
PDFs and their evolution. 
The direct calculation of PDFs using lattice QCD poses particular challenges 
due to the Euclidean geometry and the light-cone dominance of the kinematics.
The connection between PDFs and 
hadronic matrix elements, which are calculable in lattice QCD, is 
established through the 
moments of the PDFs, denoted as 
$\left\langle x^{n} \right\rangle$~\cite{Curci:1980uw,Collins:1981uw}.
Lattice QCD calculations of the moments of the PDFs, 
pioneered in Refs.~\cite{Kronfeld:1984zv,Martinelli:1987zd,Martinelli:1987bh}, 
provide, in principle, a means for the 
complete reconstruction of the PDFs. 
This possibility has remained impractical 
due to the theoretical and numerical challenges 
associated with computing high moments.
Alternative approaches have been developed to determine 
the $x$-dependence of the PDFs, the heavy-quarks OPE~\cite{Aglietti:1998mz,Detmold:2005gg}, 
the quasi-PDF~\cite{Ji:2013dva},  
the psuedo-PDF~\cite{Radyushkin:2016hsy,Karpie:2018zaz},
the OPE-based method of~\cite{Chambers:2017dov},
the current-current approach~\cite{Braun:2007wv,Ma:2014jla}, 
and the hadron tensor method of~\cite{Liang:2019frk}.
It is worth noticing that these approaches allow 
in principle an indirect determination of the 
moments of PDF of nucleons
~\cite{HadStruc:2021qdf,Gao:2022uhg} and pions~\cite{Gao:2020ito,Gao:2022iex}.

Different high-energy scatterings are associated with different PDFs:
unpolarized, helicity, and transversity.
Certain kinematical regions, as well as specific PDFs, 
pose challenges for extraction from experimental data. 
This emphasizes the significance
of reconstructing the PDFs from lattice QCD simulation data.

In this work, we revisit the possibility of calculating moments of PDFs, 
which, if successful, would provide an important and complementary 
approach to determine distribution functions from QCD.
We describe a method that addresses both the theoretical and numerical 
challenges faced in the past, which hindered the calculation of 
moments of any order from lattice QCD.

{\it{Twist-2 operators.}} 
QCD factorization relates the moments of the structure functions
and the moments of the parton distribution functions.
Taking the example of $F_1$, the relation is given by 
e.g.
\be 
\int_0^1 dx~x^{n-1} F_1(x,Q^2) = \sum_a C_{1,n}^a(Q^2/\mu^2)
A_n^{a/h}(\mu)
\label{eq:def_mom}
\ee 
where $A_n^{a/h}(\mu)$ represents the reduced matrix element
of certain local operators, denoted by $O_n$, between the hadron states, denoted as $h$, and 
$C_{1,n}^a(Q^2/\mu^2)$ are the corresponding Wilson coefficients, calculable 
in perturbation theory.
In the context of DIS kinematics, 
the leading contribution, given in Eq.~\eqref{eq:def_mom}, 
arises from local operators
with the lowest twist, $\tau = d_O - n = 2$, where $d_O$ is the physical 
dimension of the local operator and $n$ is its spin.
In the parton model the reduced matrix element is related to the moments
of the PDFs, $A_n^{a/h} = \left\langle x^{n-1} \right \rangle_{a/h}$~\cite{Curci:1980uw,Collins:1981uw}.

The form of the local operators $O_n$ can be constrained 
by requiring that they belong 
to irreducible representations of the Lorentz group.
This guarantees that the renormalization procedure does not generate 
mixing with operators of the same or lower dimensions.
Lattice QCD simulations are performed in a Euclidean setup though, and
the symmetry group in a Euclidean lattice is the hyper-cubic group H($4$).
The reduced symmetry does not protect the operators from a complicated mixing pattern,
that gets more complicated for larger $n$.
Once the continuum limit has been performed, one recovers O($4$) symmetry.

The construction of irreducible representations of O($4$) symmetry group 
is standard material (see, for example, Ref.~\cite{hamermesh1962group}).
For the purpose of this work, the only relevant result is that 
traceless and symmetrized rank-$n$ tensors are an irreducible 
representation of O($4$). 

We restrict ourselves to unpolarized structure functions, 
meaning that we consider hadronic matrix elements averaged 
over the hadron polarization. 
Other moments of different distributions can be studied 
using the same procedure described here. 
We further limit our consideration to flavor non-singlet 
operators to avoid mixing with gluonic operators,
but the same method is applicable to singlet operators.

We consider the following set of operators
\be 
O_n^{rs}(x) = O^{rs}_{\mu_1 \cdots \mu_n}(x) =
\psibar^r(x) \gmuopen1 \lrDmu2 \cdots \lrDmuclose{n} \psi^s(x)\,,
\label{eq:t2}
\ee 
with quark fields, $\psi^r$, of different flavors, $r \neq s$, and the 
covariant derivative $D_\mu = \partial_\mu + G_\mu$, where $G_\mu$ 
represent the gluon field.
The symmetrization of the covariant derivative 
$\lrD_\mu = \frac{1}{2}\left({\overset\rightarrow{D}_\mu} - 
{\overset\leftarrow{D}_\mu}\right)$
guarantees a definite transformation under charge conjugation.

In the continuum,
traceless non-singlet twist-2 operators, with symmetrized 
Lorentz indices,~\footnote{We denote the traceless operators with $\widehat{O}_n$.} 
renormalize multiplicatively.
For example, in the $\MS$ scheme, the bare operator $\widehat{O}_{n,{\text{B}}}^{rs}$ 
renormalizes as follows
\be 
\widehat{O}_n^{rs}(x) = Z^{\MS}_n \widehat{O}_{n,{\text{B}}}^{rs}(x)\,,
\ee 
with renormalization constant $Z^{\MS}_n$.
The intrinsic non-perturbative nature of the hadronic matrix elements 
suggests the use of lattice QCD. However, the breaking 
of the rotational group symmetry into the hypercubic group H($4$)
adds complications to the renormalization. 

On an hypercubic lattice, the Lorentz indices of the 
twist-2 operators must be chosen while considering the reduced
hypercubic symmetry group H($4$). 
Irreducible representations of O($4$) generally become reducible 
representations of H($4$) inducing 
unwanted mixings under 
renormalization~\cite{Kronfeld:1984zv,Martinelli:1987zd,Beccarini:1995iv,Gockeler:1996mu}.
The irreducible representations of H($4$) 
allow mixings with lower-dimensional operators,
and, additionally, complicate the mixing with operators of the same dimensions.
A notable example of this complication occurs with the operator $O_3$. 
In the continuum, after
symmetrizing over the Lorentz indices and subtracting the traces, 
the renormalization is multiplicative (e.g. in the $\MS$ scheme). 
However, on the lattice, terms like $1/a^2 \delta_{\mu_i \mu_j} \cos( a p_{\mu_j})$ 
are allowed by H($4$) symmetry and cannot be removed even after subtracting 
the trace~\cite{Kronfeld:1984zv}.
This pertains to the fact that $O_3$ belongs 
to different irreducible representations of H($4$) 
depending on the choice of the Lorentz indices.
The operator $O_3$ with 
all the indices equal, i.e. $\mu_1 = \mu_2 = \mu_3$, mixes with the 
lower-dimensional vector current, causing a power divergence proportional 
to $1/a^2$. For the operators with $\mu_1 \neq \mu_2 = \mu_3$,
such as the irreducible representation $O_{411} - O_{433}$,
the renormalization is multiplicative, up to a small mixing with 
a different operator of the same dimension.\footnote{
    The operator $O_{411} - O_{433}$ belongs to an irreducible 
    representation of H($4$) that appears more than once. 
    Therefore, a mixing between operators
    belonging to these equivalent representations can occur~\cite{Beccarini:1995iv}.
    In perturbation theory the mixing seems to be numerically small.
}
We note that while $O_{411}$ is also affected by power divergences,
the subtraction $O_{411} - O_{433}$ 
guarantees their removal.
Another choice is $\mu_1 \neq \mu_2 \neq \mu_3$. In this case, the operator belongs 
to another irreducible representation of H($4$) and renormalizes multiplicatively 
without introducing power divergences.
However, this last example, while optimal in terms of renormalization, is far from 
optimal when calculating hadronic matrix elements\footnote{In Eq.~\eqref{eq:matrix} for simplicity 
we omit the flavor indices.}
\be 
\left \langle h(p)| \widehat{O}_n |h(p) \right\rangle\ = 
p_{\mu_1} \cdots p_{\mu_n} A_n^{h}(\mu)\,,
\label{eq:matrix}
\ee 
because it requires nonvanishing external momentum in at least $2$ spatial directions.
This substantially degrades the quality of the signal-to-noise ratio
that is realized in numerical lattice QCD simulations.
For $O_{411} - O_{433}$, one needs a nonvanishing external momentum in at least 
$1$ spatial directions, and despite the subtraction of power divergences, 
one might hope to achieve a better signal-to-noise ratio.
The operator with all the same indices has a power divergence that requires 
a complicated nonperturbative procedure to be subtracted and is 
not a viable solution. This is unfortunate because the matrix element 
of $O_{444}$, for example, does not require any external spatial momentum
and should be optimal in terms of the signal-to-noise ratio.

The mixing pattern becomes more cumbersome for higher dimensional 
operators, $n \ge 4$. For $n=4$, choosing 
all $4$ indices differently results in the only $2$ irreducible 
representations: the totally symmetric and antisymmetric representations.
One then needs a nonvanishing external momentum in all $3$ spatial directions
to calculate the hadronic matrix elements~\cite{Beccarini:1995iv}. 
For $n > 4$, mixings are unavoidable. 
The complicated renormalization pattern just described, 
together with the need of an external momentum to calculate 
the hadronic matrix elements, has, {\it de facto}, prevented the lattice calculation 
of the moments of parton distribution functions for $n>4$ and rendered the calculations
for $n=3,4$ challenging.

A very interesting idea to circumvent the renormalization issue, 
based on the recovery of the continuum rotational symmetry, 
was proposed in Ref.~\cite{Davoudi:2012ya}. 
Even though the method was not further pursued numerically, 
it represents a first important step toward resolving 
the theoretical challenges encountered when calculating moments of PDFs.

{\it{Gradient flow.}} 
The gradient flow (GF) for gauge and fermion 
fields~\cite{Luscher:2010iy,Luscher:2011bx,Luscher:2013cpa} provides 
a different way to regulate short-distance singularities 
of the twist-2 operators.
The connection with the physical renormalized matrix elements at vanishing flow time,
$t=0$, is obtained with a short-flow-time expansion (SFTX) after performing
the continuum limit of the hadronic matrix elements at a fixed 
physical value of the flow time $t$~\cite{Luscher:2013vga}.

The flowed twist-2 operators, written in terms of flowed fermions, $\chi^r(x,t)$ 
and $\chibar^r(x,t)$, and gauge fields, $B_\mu(x,t)$, are given by 
\be 
O_n^{rs}(x,t) = \chibar^r(x,t) \gmuopen1 \lrDmu2 \cdots \lrDmuclose{n} \chi^s(x,t)\,,
\label{eq:flowed_t2}
\ee 
where $\left\{\cdots\right\}$ denotes symmetrization over the included indices.
These operators are particularly advantageous because they renormalize
multiplicatively also on the lattice with a renormalization 
factor that depends only on the fermion content of the operator,
\be 
O_n^{rs}(t) = Z_n O_{n,B}^{rs}(t)\,, \quad Z_n = Z_\chi\,,
\ee 
where $Z_\chi^{1/2}$ is the renormalization constant of the flowed 
fermion fields~\cite{Luscher:2013cpa}. 
For correlation functions containing flowed twist-2 operators, 
the flow-time $t$ provides a regulator for short-distance singularities. 
Beside the renormalization of the bare parameters of the theory,
the flowed fermion fields, and any local field at $t=0$, 
correlation functions do not 
require any additional renormalization.
For lattice QCD applications, it is convenient to define $Z_\chi$
in a scheme that is regularization independent, as then 
the matching and the renormalization can be performed in the same scheme.
The choice of the specific scheme is not relevant for the general discussion,
but it becomes important when calculating the matching coefficients.
In the calculations of the matching coefficients described below, 
we renormalize the twist-2 operators using the so-called ringed fields~\cite{Makino:2014taa},
defined by the SU($N_c$) gauge invariant and regularization independent condition 
\be 
\left \langle \rchibar_r(x,t) {\overset\leftrightarrow{\slashed{D}}} \rchi_r(x,t) \right\rangle
= - \frac{N_c}{(4 \pi)^2 t^2}\,.
\ee 
Imposing this condition in dimensional regularization, with $D=4 - 2 \epsilon$,
leads to a finite renormalization between the ringed fields and 
renormalized fields in the $\MS$ scheme
\bea 
\chi_r(x,t) &=& \left(8 \pi t\right)^{\epsilon/2} \zeta_\chi^{1/2} \rchi_r(x,t) \nonumber \\ 
\chibar_r(x,t) &=& \left(8 \pi t\right)^{\epsilon/2} \zeta_\chi^{1/2} \rchibar_r(x,t)\,.
\eea 
The 1-loop result for the finite renormalization is given by~\cite{Makino:2014taa} 
\be 
\zeta_\chi = 1 - \frac{\gbar^2(\mu)}{(4 \pi)^2}C_F 
\left( 3 \log (8 \pi \mu^2 t) - \log (432)\right)\,,
\ee 
where $\gbar$ is the strong coupling renormalized in the $\MS$ scheme
at the renormalization scale $\mu$, and $C_F = N_c^2-1/2N_c$. 
This result has been extended to O($\gbar^4$) in 
Refs.~\cite{Harlander:2018zpi,Artz:2019bpr}.
Results obtained in the $\MS$ scheme can be converted to the 
$\MSbar$ scheme by replacing the renormalization scale $\mu$ with 
$\mubar$ related by $\log \mu^2 = \log \mubar^2 + \gamma_E - \log 4 \pi$.

{\it{O($a$) improvement.}} 
The lattice QCD calculation of hadronic matrix elements of twist-2 operators 
is affected by cutoff effects. 

For nonperturbatively improved clover fermions, 
the hadronic matrix elements of the standard 
twist-2 operators (with $t=0$) are still affected 
by O($a$) cutoff effects that could be removed by
adopting an O($a$) improved definition of the twist-2 operators.
Excluding the case $n=2$, where the 
corresponding improvement coefficients have been determined 
at 1-loop in perturbation theory~\cite{Capitani:2000xi},  
the continuum limit is reached with O($a$) 
discretization effects.

If we consider flowed twist-2 operators for a generic $n$, 
the hadronic matrix elements 
are affected only by O($a m$) cutoff effects, 
where $m$ is the quark mass. Additional improvement 
coefficients specific to the twist-2 operators are not needed.
In the discussion of the method's applications
below, we also consider 
ratios of flowed matrix elements where both $Z_\chi$ and 
the O($am$) cutoff effects are eliminated. 
In summary, regardless of the lattice setup adopted, 
the use of flowed twist-2 operators not only provides a significant simplification 
in the renormalization pattern, but also 
improves the parametric scaling in the lattice spacing
towards the continuum limit.

{\it{Matching coefficients.}}
After performing the continuum limit of the hadronic matrix element
for a fixed $t>0$ value of the flow time, the connection with 
physical renormalized matrix element at $t=0$ is obtained 
with a SFTX~\cite{Luscher:2013vga}.
The SFTX of the twist-2 operators $O_n(x,t)$, in general contains power 
divergences with terms like $1/t^{m} \delta_{\mu_i \mu_j}$,
where $m=(d_n - d_p)/2$ depends on the dimensions of $O_n$, $d_n$,
and on the dimension of the lower-dimensional operator, $d_p$. 
These terms, though, are classified using the continuum O($4$) symmetry and, 
as a result, the SFTX of the symmetrized and traceless operators,
$\widehat{O}_n$, receives 
contributions only from the corresponding traceless operators at $t=0$.

We renormalize the flowed fermion fields adopting 
ringed fields, introduced in the previous section, and define the ringed twist-2
operators
\bea 
\widehat{\rO}_n^{rs}(x,t) &=& 
\rchibar^r(x,t) \gmuopen1 \lrDmu2 \cdots \lrDmuclose{n} \rchi^s(x,t) - \nonumber \\ 
&-& \text{terms~with~}\delta_{\mu_i \mu_j}\,,
\label{eq:flowed_t2}
\eea 
where the subtraction guarantees that the resulting operator has vanishing traces.
The SFTX of $\widehat{\rO}_n^{rs}$ is 
\be 
\widehat{\rO}_n^{rs}(t) = c_n(t,\mu)\widehat{O}_{n,\MS}^{rs}(\mu) + \cdots\,,
\ee 
where we make explicit only the dependence on the flow time and 
the renormalization scale. The neglected terms are higher-order
contributions of positive powers of $t$.

The matching coefficients are determined considering 
off-shell amputated one-particle irreducible (1PI) Green's functions
containing the flowed operators $\widehat{\rO}_n^{rs}(t)$.
The matching equations with massless fermions
\be 
\left\langle \psi^r \widehat{\rO}_n^{rs}(t) \psibar^s \right\rangle = c_n(t,\mu)
\left\langle \psi^r \widehat{O}_{n,\MS}^{rs}(t=0,\mu) \psibar^s \right\rangle
\label{eq:matching_eq}
\ee 
are solved in $D = 4 - 2 \epsilon$ at 1-loop in perturbation theory with
\be 
c_n(t,\mu) =  1 + \frac{\gbar^2(\mu)}{\left(4 \pi\right)^2}c_n^{(1)}(t,\mu) + 
O(\gbar^4)\,.
\label{eq:cn_1l}
\ee 
The matching coefficients $c_n(t,\mu)$ depend on the renormalized coupling 
and are independent of the soft scales the calculation of the matching coefficients.
In the Supplemental Material~\cite{SupplMat} we describe their calculation 
and, based on the observation that they satisfy a renormalization 
group equation, we also provide a resummed expression at next-to-leading log (NLL).

The final result for the matching coefficients is 
\be 
c_n^{(1)}(t,\mu) = C_F \left[ \gamma_n \log \left(8 \pi \mu^2 t \right) + B_n\right]\,,
\label{eq:matching_1l}
\ee 
where $\gamma_n = 1 + 4 \sum_{j=2}^n \frac{1}{j} - \frac{2}{n(n+1)}$, and
\bea 
B_n &=& \frac{4}{n(n+1)} + 4 \frac{n-1}{n}\log 2 + \frac{2-4 n^2}{n(n+1)}\gamma_E - \nonumber \\ 
&-& \frac{2}{n(n+1)}\psi(n+2) + \frac{4}{n}\psi(n+1) - 4 \psi(2) - \nonumber \\
&-& 4 \sum_{j=2}^n \frac{1}{j(j-1)} \frac{1}{2^j} \phi(1/2,1,j) - \log \left(432\right)\,.
\label{eq:finite}
\eea 
In this expression $\gamma_E = 0.57221\ldots$ is the Euler's constant, $\psi(z)$
is the digamma function, and $\phi(z,s,a)$ is the Lerch transcendent defined as 
$\phi(z,s,a) = \sum_{k=0}^{\infty} \frac{z^k}{(k+a)^s}$.

The expression for $\gamma_n$ matches the 1-loop anomalous dimension of the twist-2
operators~\cite{Gross:1974cs} and provides a welcome check of the calculation.
For $n>2$ the finite part, $B_n$, in Eq.~\eqref{eq:finite} provides a new result. 
The result with $n=2$ was already obtained 
in Ref.~\cite{Makino:2014taa} in the analysis of the energy-momentum tensor, 
and we reproduce their result.

The fact that the matching of the flowed matrix elements is so simple
opens the possibility of computing moments of the parton distribution functions 
of any order.
A possible operational procedure is the following: After fixing $n$,
one constructs the symmetric and traceless operator $\widehat{O}_n^{rs}$.
One proceeds calculating the matrix elements between hadron states of 
$\widehat{\rO}_n^{rs}(t)$
using standard techniques based on the spectral decomposition.
To apply the decomposition 
in the lattice calculation of the 3-point function, 
one needs to ensure that the physical distance 
between the interpolators of the 
hadron states and of the flowed operators is much larger than 
the flow-time radius $\sqrt{8t}$. After performing the continuum limit, 
the renormalized matrix element in the $\MS$ scheme is calculated 
simply by 
\be 
A_n^{\MS}(\mu) = 
c_n(t,\mu)^{-1} A_n(t) \,.
\label{eq:x_MS}
\ee 
This matching is correct only if the flowed fields are defined as ringed fields.
If the flowed fields are renormalized in a different scheme, the finite 
part of the matching coefficient has to be modified.

If the matching is successful, the l.h.s of Eq.~\eqref{eq:x_MS} 
should be independent of the flow time.
Violations stem from higher dimensional operators as linear terms in $t$,
and from higher order terms in the perturbative 
expansion of the matching coefficients.
To mitigate these possible systematics it is important to 
extend to O($\gbar^4$) the calculation presented here
and conduct a thorough numerical study of the residual
flow-time dependence of the hadronic matrix element after the matching.

{\it{Applications.}} 
The simplicity of the matching is guaranteed once the operators are symmetrized
on the Lorentz indices, and all the traces have been subtracted.
Traceless operators for any $n$ are not difficult to write down.
For example, for $n=4$, we have 
\be 
\widehat{O}_{4444} = O_{4444} - \frac{3}{4}O_{\left\{\alpha \alpha 4 4 \right\}} 
+ \frac{1}{16} O_{\left\{\alpha \alpha \beta \beta \right\}}\,.
\ee 
The hadronic matrix element of $\widehat{O}_{4444}(t)$ is calculable 
with no external spatial momentum, and the matching with the 
$\MS$ scheme at renormalization scale $\mu$ is obtained
using Eq.~\eqref{eq:x_MS}.

The continuum limit of the reduced hadronic matrix element of $\widehat{\rO}_n$,
denoted as $A_n^h(t) = \left\langle x^{n-1} \right\rangle^h(t)$, 
would require a determination of $Z_\chi$ from lattice QCD simulations.
However, this can be avoided using the second moment 
$A_2^{h,\MS}(\mu) = \left\langle x \right\rangle_{\MS}^h(\mu)$ as an 
observable input already computed 
with any method of choice. Then all the other moments for $n>2$ 
are calculable from 
\be 
\left\langle x^{n-1} \right\rangle_{\MS}^h(\mu) = 
\left\langle x^{n-2} \right\rangle_{\MS}^h(\mu)
\frac{c_{n-1}(t,\mu)}{c_n(t,\mu)} R_n^h(t)\,,
\label{eq:ratio_n2}
\ee 
where $c_n(t,\mu)$ is given by 
Eqs.~(\ref{eq:cn_1l}-\ref{eq:finite}). 
The ratios 
\be 
R_n^h(t) = \frac{\left\langle x^{n-1} \right\rangle^{h}(t)}{\left\langle x^{n-2}\right\rangle^{h}(t)}\,,
\qquad n>2\,,
\label{eq:Rn}
\ee 
have a finite continuum limit, 
and there is no need to renormalize the flowed fields.
Moreover, these ratios exhibit O($a^2$) scaling violations 
once the lattice theory has been nonpertubatively improved.
If convenient numerically, it is also possible to use different hadrons 
for the calculation of the moment matrix elements in 
Eq.~\eqref{eq:ratio_n2}
because the matching coefficients are independent on the hadron 
for which we calculate the matrix element.

The method proposed in this paper can be 
also generalized to study distribution amplitudes,
but care should be taken when studying the 
short flow time expansion due to the presence of operators
with vanishing contributions for forward matrix 
elements~\cite{Bali:2017ude,RQCD:2019osh}.

The gradient flow has previously been employed to propose a locally smeared 
operator product expansion in scalar field theory~\cite{Monahan:2015lha} 
and to define quasi-PDF in~\cite{Monahan:2016bvm}, 
with the 1-loop matching calculated in~\cite{Monahan:2017hpu}.
For a more realistic application of the gradient flow 
in quasi-PDF calculations see Ref.~\cite{HadStruc:2022yaw}.
While, in principle, the determination of moments of PDFs
should allow a precise reconstruction of the PDFs,
it would be intriguing to explore whether the determination of the 
PDF moments outlined in this work 
can be combined with quasi-PDF calculations.

{\it{Summary.}} 
We have described a method that resolves both the theoretical 
and numerical challenges faced when calculating moments 
of parton distribution functions from lattice QCD.
A lattice regulator breaks the 
Euclidean rotational symmetry into the hypercubic group.
As a result, the renormalization of twist-2 operators 
on the lattice is a substantial hurdle to determine 
moments of parton distribution functions 
$\left\langle x^{n-1} \right\rangle$.
While for $n>4$, power divergences in the 
lattice spacing are unavoidable, for $n=3,4$ 
the cancellation of power divergences is still possible
by choosing suitable irreducible representations 
of the hypercubic group. The price to pay is that 
the calculation of the hadronic matrix elements 
necessitates the use 
of a spatial external momentum, worsening the signal-to-noise 
ratio of the correlation functions. 
At finite flow time $t$, the correlation functions 
of twist-2 operators
are finite once the bare parameters of the theory and the flowed fermion 
fields are renormalized. 
After performing the continuum limit, 
the matching with the physical matrix element can be done 
at short flow time using the continuum symmetries of the theory.
Using traceless and symmetrized operators guarantees that the 
matching is multiplicative, and we have calculated, at 
1-loop in perturbation theory, the matching coefficients.

It would be also very interesting 
to investigate how the approach described in this work 
can complement recent efforts aimed at the direct 
determination of PDFs from lattice QCD correlators.

While, the method proposed in this paper is applicable, 
at least in principle, to calculate any moments 
of parton distribution functions,
in practice lattice QCD calculations of hadronic matrix elements of 
the flowed twist-2 operators in Eq.~\eqref{eq:flowed_t2} are subject to 
several systematic uncertainties.

In the Supplemental Material~\cite{SupplMat} there is a discussion of the
systematics, but only a thorough numerical investigation 
provides the final response about the 
range of applicability, and whether 
this result opens the way to determine moments of any order 
of the parton distribution functions.

{\it{Acknowledgments}}
I would like to thank Andre Walker-Loud
for constant encouragement and a critical reading of the manuscript. 
I also thank R. Harlander and T. Luu for suggestions 
on how to improve the original draft. 
I have benefited from many discussions about the gradient 
flow with R. Harlander, J. Kim, T. Luu, E. Mereghetti, C. Monahan, M. Rizik, 
P. Stoffer and A. Walker-Loud. 
I want to thank the Nuclear Theory Group of the Lawrence Berkeley National Lab
and the University of California Berkeley for their hospitality.
I acknowledge funding support from 
Deutsche Forschungsgemeinschaft (DFG, German Research Foundation) 
through grant 513989149 and under 
the National Science Foundation Grant No. PHY-2209185.

\bibliography{refs}

\newpage\hbox{}\thispagestyle{empty}\newpage

\section{Supplemental Material}  

The method proposed in this paper is applicable, 
at least in principle, to calculate any moments 
of parton distribution functions, resolving the 
long-standing problem of renormalizing twist-2 operators 
on the lattice.
In practice lattice QCD calculations of hadronic matrix elements of 
the flowed twist-2 operators in Eq.~\eqref{eq:flowed_t2} are subject to 
several systematic uncertainties.

{\it{Finite volume effects.}}
In the continuum limit the space-time extension of the flowed 
twist-2 operators in Eq.~\eqref{eq:flowed_t2} is governed by the 
flow-time radius $\sqrt{8t}$, and the finite volume 
effects of the hadronic matrix elements 
are calculable using effective field theory methods~\cite{Detmold:2005pt}.
At finite lattice spacing $a$ the finite volume provides a bound on the 
number of covariant derivatives, thus a bound on the order of the moments
that can be simulated. The operator $\widehat{\rO}_n^{rs}(x,t)$ 
contains $n-1$ derivatives and the maximal distance from $x$ reached by the gauge links 
is $(n-1)a$. 
We now consider the example of the pion matrix element of $\widehat{\rO}_n^{rs}(x,t)$
calculated on a lattice with time extent $T$. 
A reasonable estimate for the maximal extension 
of the flowed operator is $\sim T/8$. For state-of-the-art lattice QCD simulations 
this implies that up to $n \sim 10$ there should be no obstruction for the calculation 
of the hadronic matrix elements. 
For nucleon matrix elements, the signal-to-noise ratio can provide a stronger 
constraint, that we address below.

{\it{Discretization errors.}}
For the calculation of $\llangle x^n \rrangle (t)$ for 
values of $\sqrt{8t}$ smaller than $ \simeq n a$ one can expect 
substantial discretization errors.
For larger values of $\sqrt{8t}$ there is no reason to expect 
different discretization effects than in other similar calculations.
Additionally, it is common practice in lattice QCD calculations 
to determine specific ratios, for which some 
of the potential systematic effects are reduced.

A typical application for ratios like $R_n^h(t)$, defined in Eq.~\eqref{eq:Rn},
is to substantially reduce cutoff effects with respect to the standard twist-2 
hadronic matrix elements. 
In the case of non-perturbatively improved Wilson fermions,
this can be also confirmed from a theoretical analysis based on the 
Symanzik effective theory approach.
In Ref.~\cite{Luscher:2013cpa} it has been shown that, 
beside the usual O($a$) cutoff effects
removed with the non-perturbative O($a$) improvement of the theory, 
the only additional 
cutoff effects are of O($am$), where $m$ is the quark mass\footnote{Potentially short-distance O($a$) cutoff effects could affect the lattice correlators involving flowed fields, but they are expected to be insignificant in the calculation of the hadronic matrix elements (see Ref.~\cite{Kim:2021qae} for another example where these short-distance O($a$) cutoff effects are negligible.).}. 
These cutoff effects depend only on the fermion content of the observable 
and thus cancel out in the ratio $R_n^h(t)$.
One is left with a ratio of lattice correlators 
affected only by O($a^2$) effect.

{\it{Perturbative matching.}}
The perturbative calculation of the matching coefficients
follows standard strategies 
(see \cite{Manohar:2018aog,Mereghetti:2021nkt} and Refs. therein),
expanding the integrands of the loop integrals of the matching 
equations~\eqref{eq:matching_eq} in all soft scales keeping $t$ fixed.
The $t=0$ 1-loop contribution on the r.h.s of Eq.~\eqref{eq:matching_eq}
vanishes because the expansion leads
to scaleless, i.e. vanishing, integrals in D-dimensions.
After renormalizing the gauge coupling and the flowed fermion fields,
the left-hand side is UV finite. 
The infrared singularities stemming from the expansion in the 
soft scales, and regulated in dimensional regularization, match 
exactly the UV poles of $Z_n^{\MS}$. The finite matching coefficients 
are the result of this procedure.

\begin{figure}[h]
  \includegraphics[width=0.16\textwidth]{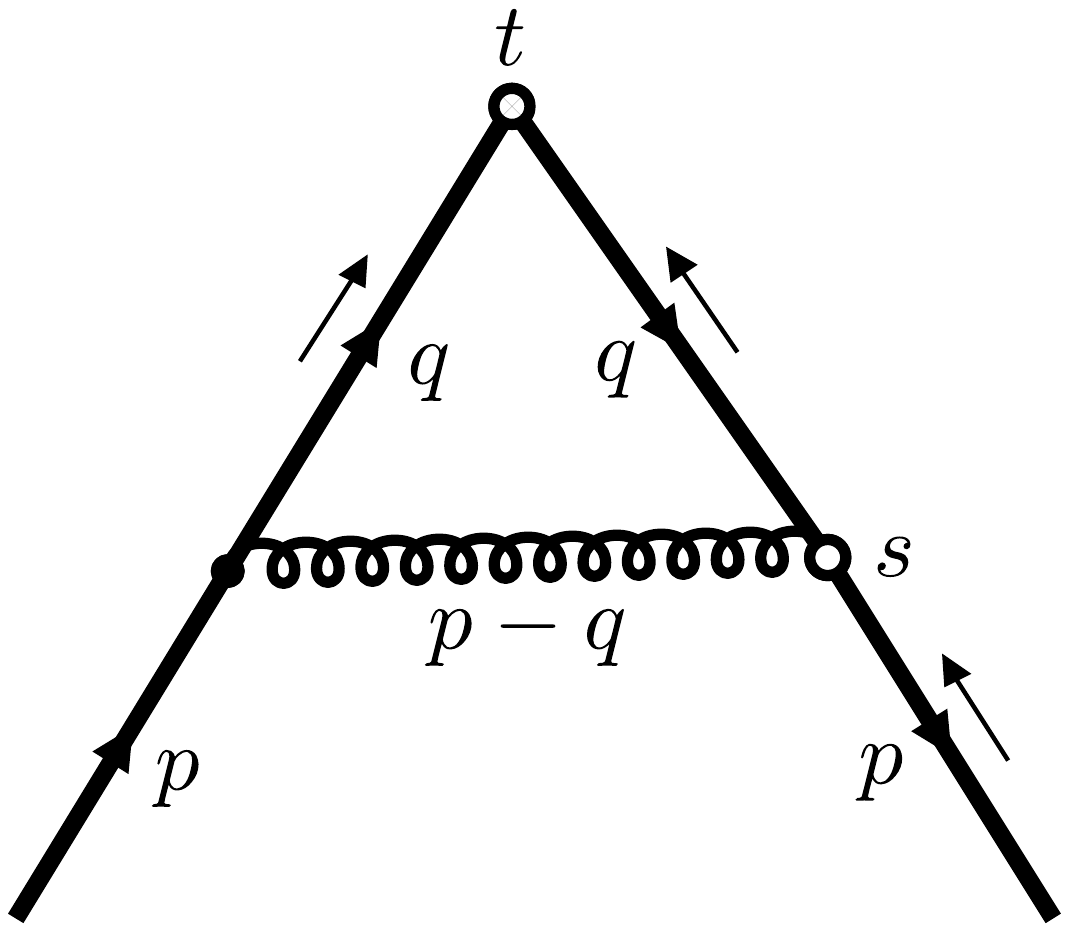}
  \includegraphics[width=0.16\textwidth]{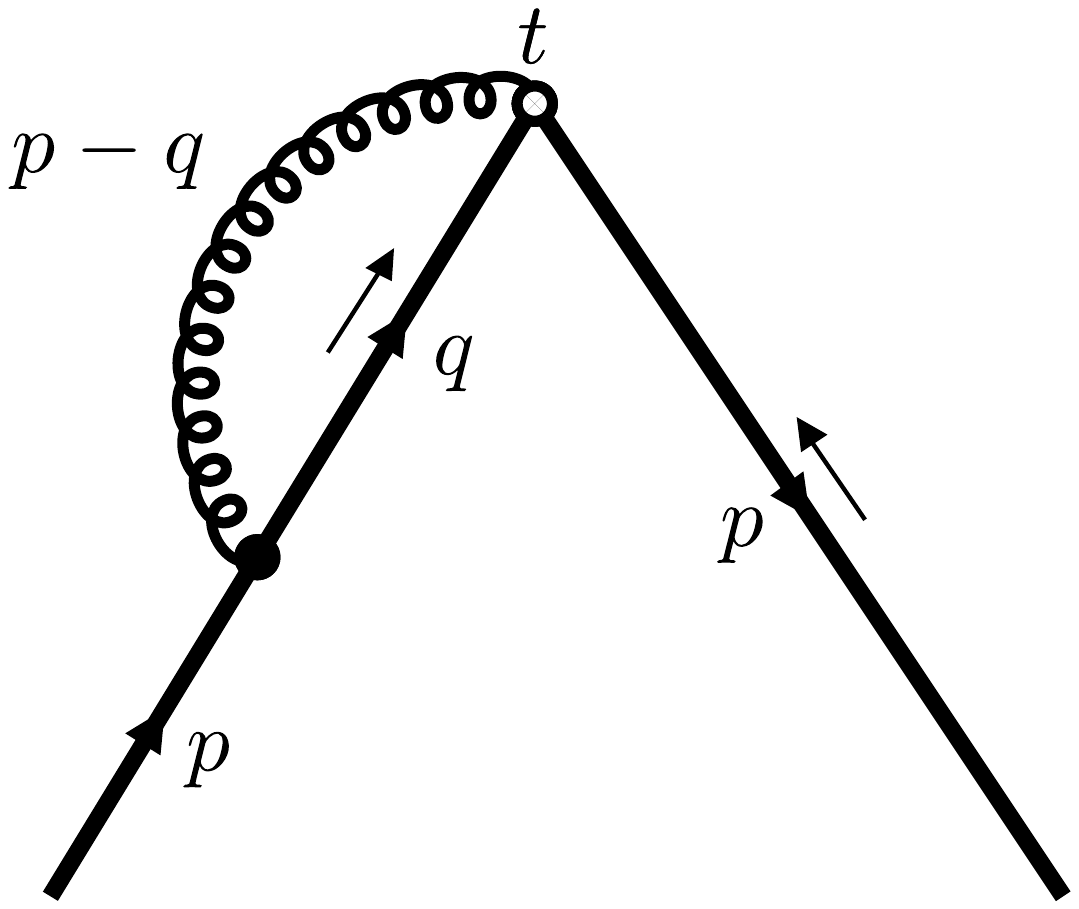}
  \includegraphics[width=0.16\textwidth]{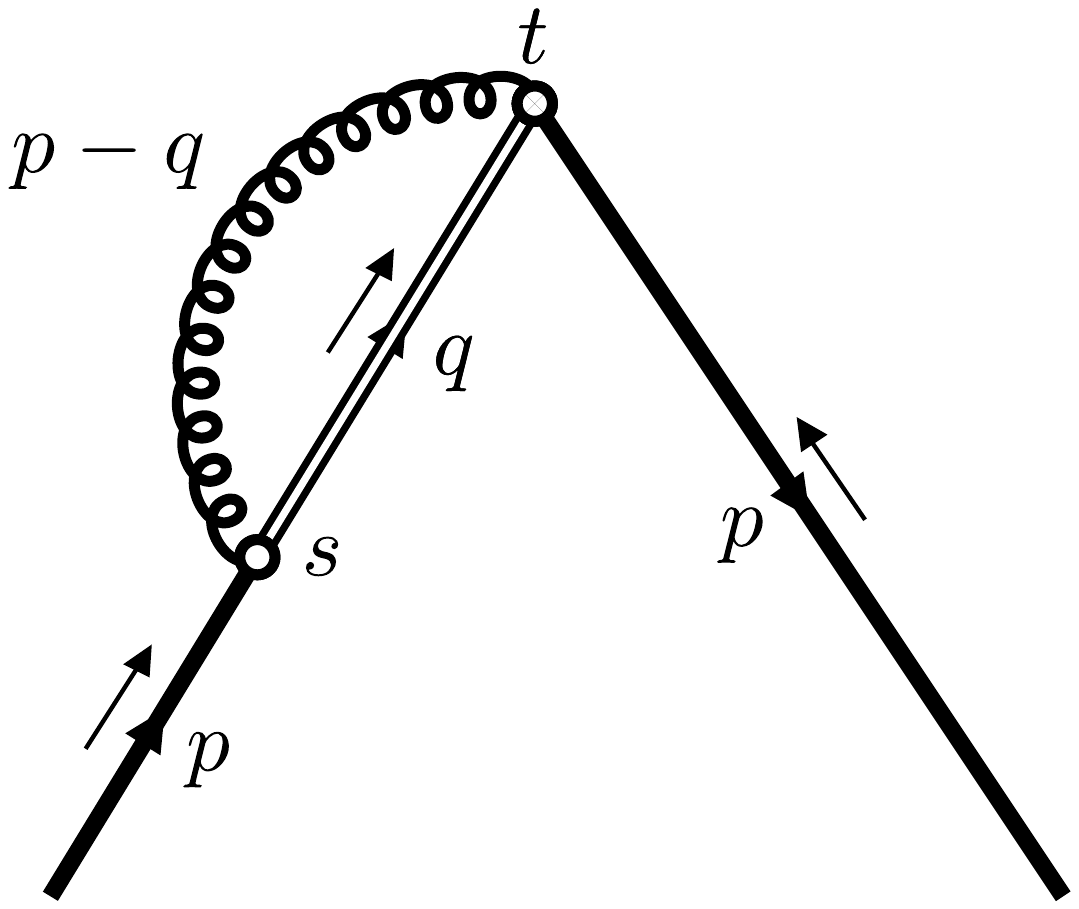}
  \includegraphics[width=0.16\textwidth]{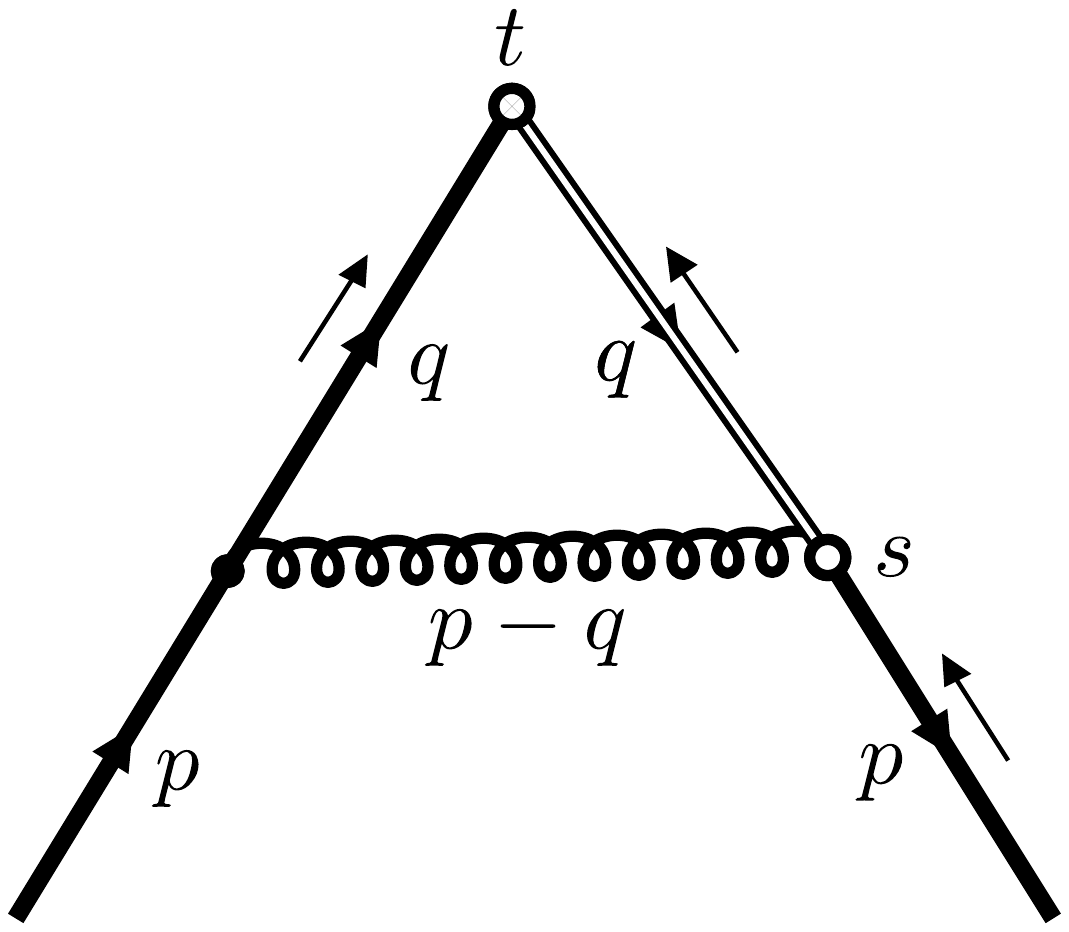}
  \includegraphics[width=0.16\textwidth]{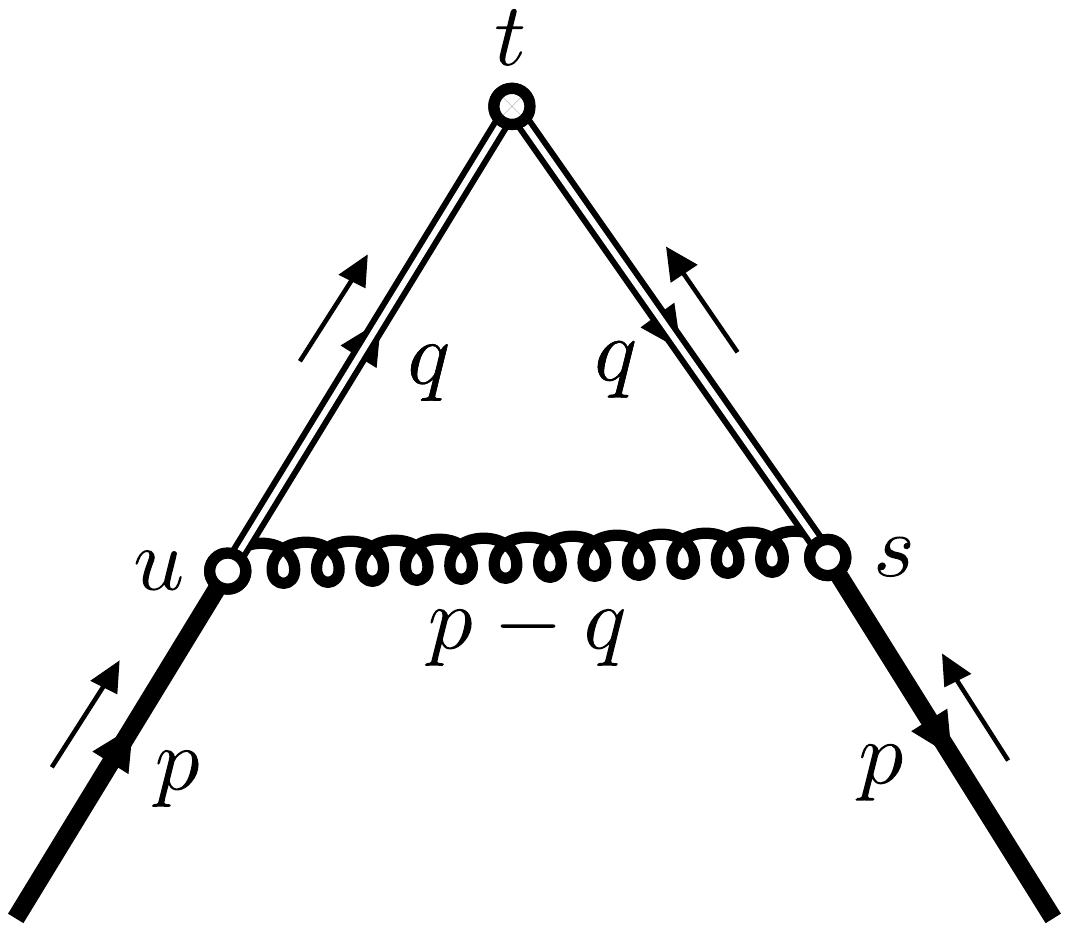}
  \includegraphics[width=0.16\textwidth]{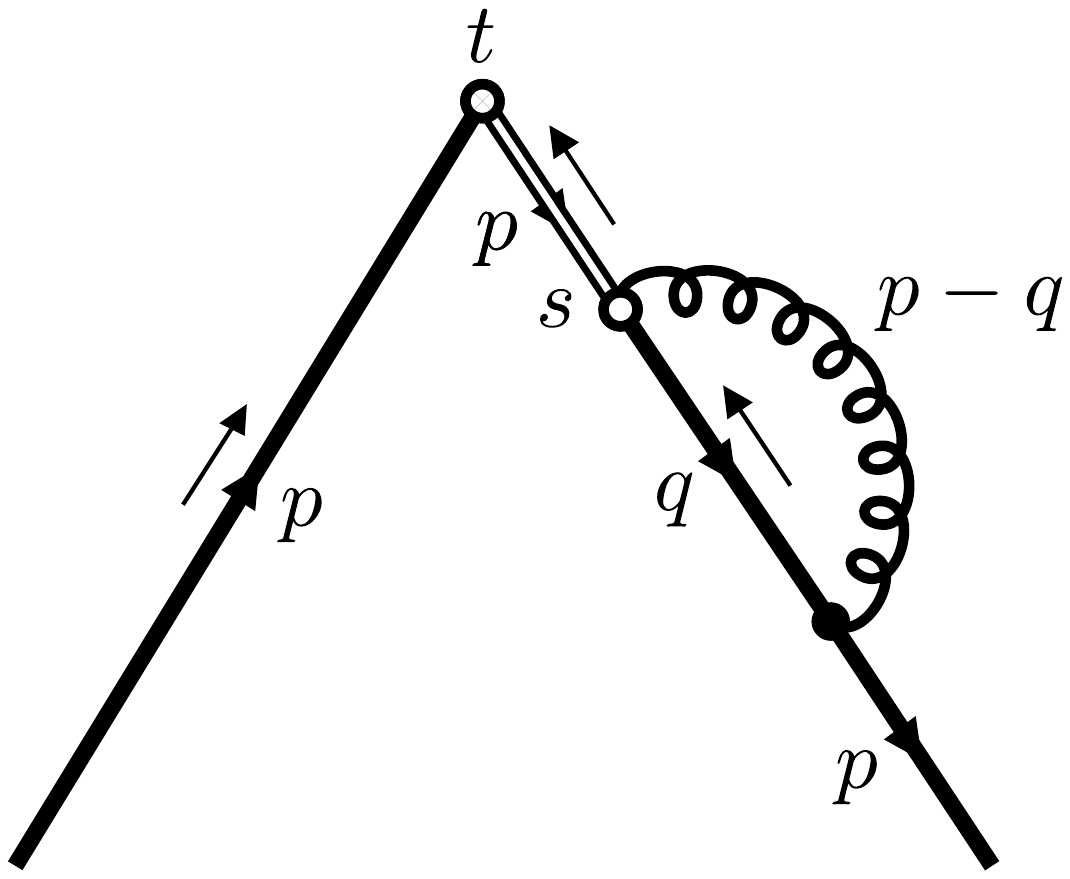}
  \includegraphics[width=0.16\textwidth]{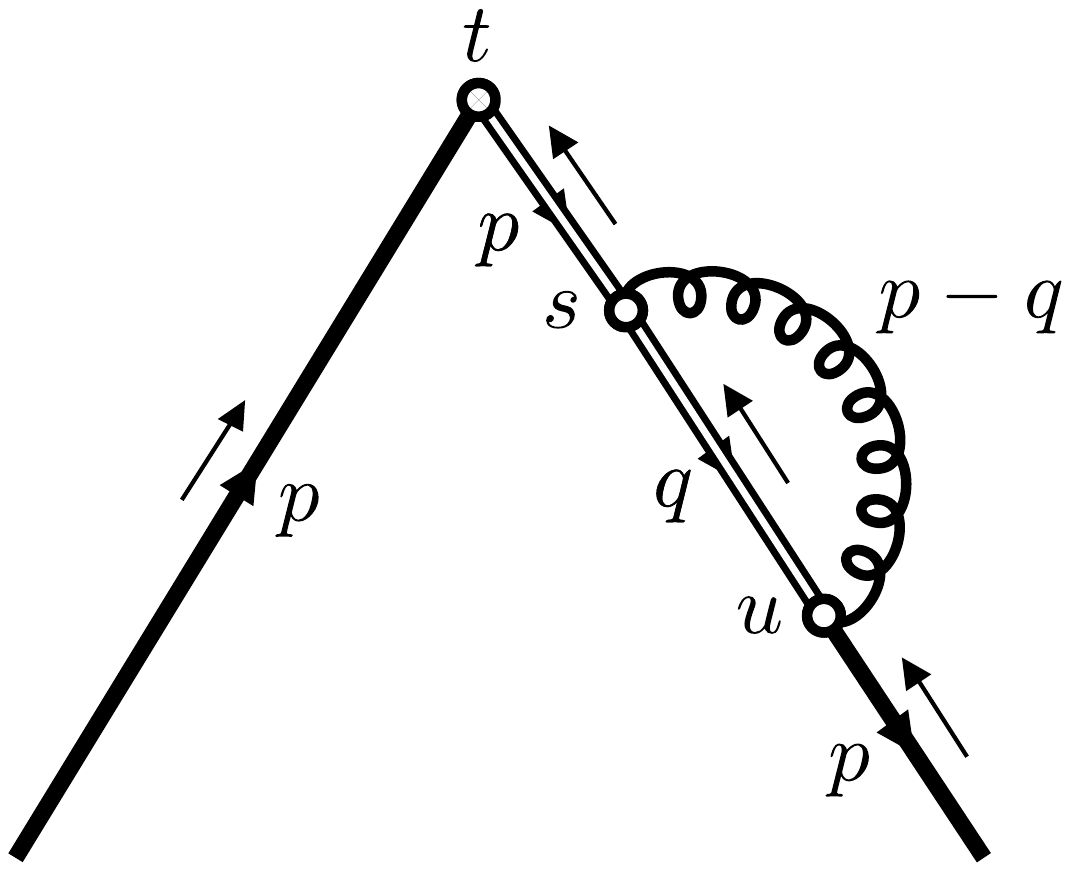}
  \includegraphics[width=0.16\textwidth]{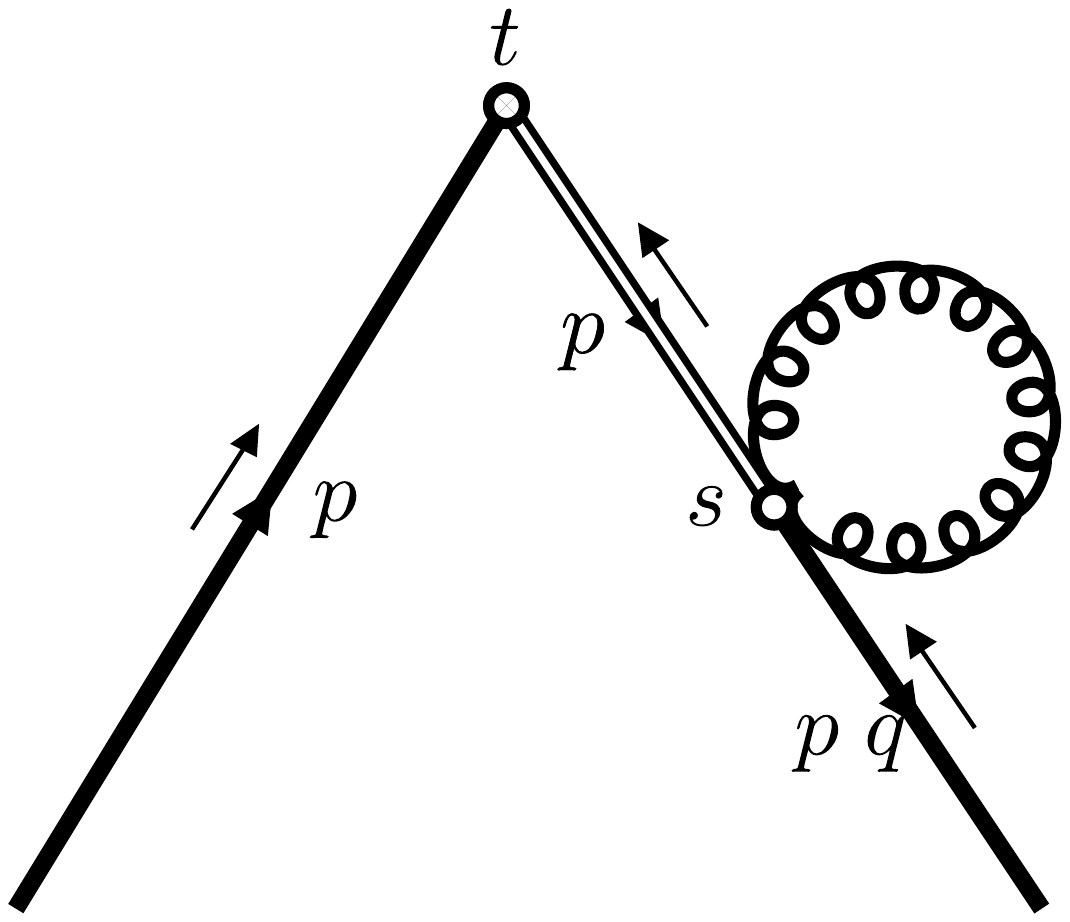}
  \caption{Feynman diagram for the O($\gbar^2$) calculations 
  of the matching coefficients in Eq.~\eqref{eq:matching_eq}. The circle with 
  the cross represents the twist-2 operator and the empty circles the GF vertices.
  The double line represents the fermionic
  kernel of the GF differential operator. 
  The arrows beside the fermion lines and the kernel lines indicate the 
  direction of increasing flow time. }
  \label{fig:fd}
\end{figure}

The Feynman diagrams contributing are 
depicted~\footnote{To draw the Feynman diagrams we have used
FeynGame~\cite{Harlander:2020cyh}.} in Fig.~\ref{fig:fd}.
The cross indicates the insertion of the twist-2 operators, 
the single straight and wavy lines respectively represent the fermion and 
gluon propagator, and the double straight lines represent the 
kernel of the fermionic GF operator.
The open circle represents a GF vertex, and the rest are 
standard QCD vertices.
The Feynman rules for the operators are the same as in the unflowed case,
and the remaining Feynman rules related to the expansion of the GF equation
can be found, for example, in Refs.~\cite{Rizik:2020naq,Mereghetti:2021nkt}.

The final result of the calculation
is given by Eqs.~\eqref{eq:matching_1l}.
The matching coefficients satisfy a renormalization group (RG) equation 
that can be determined using the fact that the "ringed" matrix elements,
or the ratios of flowed matrix elements in Eq.~\eqref{eq:ratio_n2}
are renormalization group invariant. 
The solutions of the RG equation, $c_n^{\text{RG}}(t,\mu)$, are given by
\be 
c_n^{\text{RG}}(t,\mu) = c_n(\gbar(q)) 
\text{exp}\left\{-\int_{\gbar(\mu)}^{\gbar(q)}dx \frac{\gamma_n(x)}{\beta(x)}\right\}\,,
\label{eq:cn_resum}
\ee   
where $\beta$ is the beta function and $\gamma_n$ the anomalous dimension of the 
twist-2 operators in Eq.~\eqref{eq:t2}.
The matching coefficients $c_n(\gbar(q))$ are the 
matching coefficients in Eq.~\eqref{eq:cn_1l} with $q=1/\sqrt{t}$.
We consider the perturbative expansion of $\beta(\gbar)$
up to order $\gbar^5$~\cite{Jones:1974mm},
$\gamma_n(\gbar)$
up to order $\gbar^4$~\cite{Floratos:1977au,Gonzalez-Arroyo:1979guc,Curci:1980uw} and 
evaluate the next-to-leading-log (NLL) expression in 
Eq.~\eqref{eq:cn_resum}, $c_n^{\text{NLL}}(t,\mu)$.
\begin{figure*}
\includegraphics[width=7.5cm]{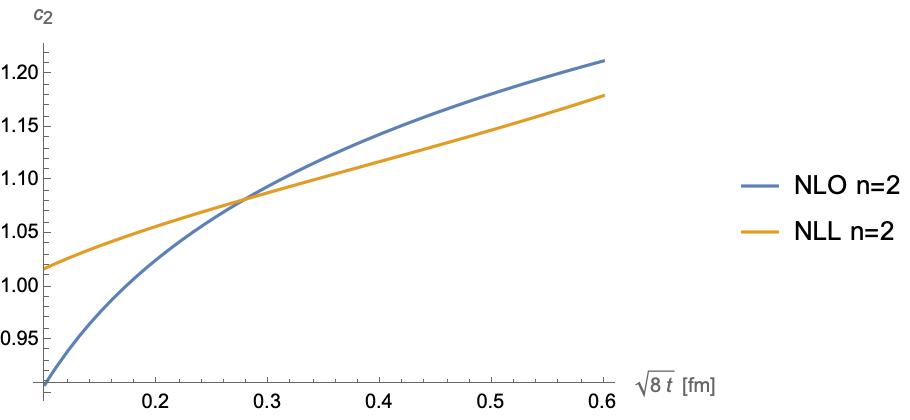}
\includegraphics[width=7.5cm]{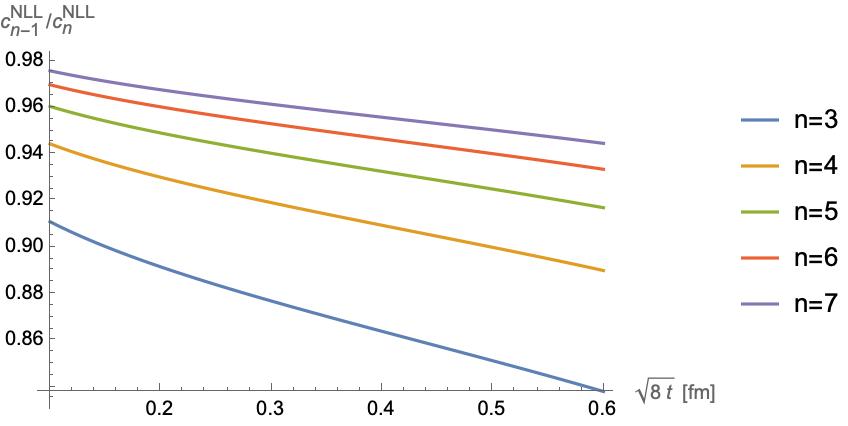}
\caption{Left Plot: flow-time dependence of the matching coefficient 
for the second moment calculated at $\mu=2$ GeV comparing the NLO result
with the resummed one.  Right Plot: flow time dependence 
of the ratios $c_{n-1}^{\text{NLL}}(\mu,t)/c_n^{\text{NLL}}(\mu,t)$ at $\mu=2$ GeV.}
\label{fig:running}
\end{figure*}
Fixing $\mu=2$ GeV 
for the second moment $\llangle x \rrangle$ the perturbative corrections 
are between $6-9\%$ in the region where the flow-time radius $\sqrt{8t}=0.2-0.3$ fm.
In the left plot of Fig.~\ref{fig:running} we show the matching 
coefficient of the second moment comparing the result
of Eq.~\eqref{eq:matching_1l} with the resummed result of Eq.~\eqref{eq:cn_resum} 
for $\mu=2$ GeV as a function of $\sqrt{8t}$. 
The range of the flow-time radius is chosen to be in the range 
where for lattice spacings 
$a=0.05-0.08$ fm large cutoff effects are not expected.
We observe that the resummation guarantees that the coupling goes smoothly 
to zero in the $t \rightarrow 0$ limit.
In a lattice QCD simulation,
after the perturbative matching, the residual flow time dependence could be removed 
extrapolating to $t \rightarrow 0$ the numerical data. 
This is a standard procedure (see for example Refs.~\cite{Black:2023vju}) 
for calculations involving flowed observables.

For higher moments we consider the ratios in Eq.~\eqref{eq:Rn}
that can be used to calculate the moments in the $\MSbar$ scheme using 
Eq.~\eqref{eq:ratio_n2}.
Fixing again the renormalization scale at $\mu=2$ GeV
in the right plot of Fig.~\ref{fig:running} we show the dependence 
of the ratios 
$\frac{c_{n-1}^{\text{NLL}}(t,\mu,\gbar(\mu))}{c_n^{\text{NLL}}(t,\mu,\gbar(\mu))}$ 
on $\sqrt{8t}$, for $n=3,\cdots, 7$. We observe that the perturbative 
corrections decrease for increasing $n$. To give a first estimate of what 
uncertainty to expect with a NLL perturbative matching,
we have calculated the perturbative corrections, $\delta_{\text{PT}}$, 
for higher moments evaluated at $\sqrt{8t} = 2 (n-1) a$,
i.e. at a distance twice as large as the farthest possible contribution 
from the twist-2 local operators. The estimate has been 
obtained at a lattice spacing of $a=0.05$ fm which is 
a lattice spacing reachable with state-of-the-art lattice QCD simulations.
The results are listed in Table~\ref{tab:pert}.

\begin{table}[h!]
  \centering
   \begin{tabular}{|c|c|c|} 
   \hline
   $n$ & $\llangle x^{n-1} \rrangle$ & $\delta_{\text{PT}}$ \\ [0.5ex] 
   \hline
   2 & $\llangle x \rrangle$     & $2\%$   \\ 
   3 & $\llangle x^{2} \rrangle$ & $11\%$  \\
   4 & $\llangle x^{3} \rrangle$ & $8\%$   \\
   5 & $\llangle x^{4} \rrangle$ & $7\%$   \\
   6 & $\llangle x^{5} \rrangle$ & $6\%$   \\
   7 & $\llangle x^{6} \rrangle$ & $6\%$   \\ [1ex] 
   \hline
   \end{tabular}
   \caption{Perturbative corrections for moments of PDF at a matching scale 
   of $\mu=2$ GeV, $\sqrt{8t} = 2(n-1) a$, and $a=0.05$ fm.}
   \label{tab:pert}
  \end{table}

From this exercise one can then conclude that the flow time dependence 
of the ratio $R_n^h(t)$ is very small, at least at NLL of perturbation theory,
and this is true also in the region of $\sqrt{8t}$ where lattice discretization 
effects should be under control.
After the perturbative matching is performed, 
it is conceivable that higher dimensional operators 
will contribute to the flow time dependence
of the matrix element. This has been observed practically in all calculations 
of matrix elements of local flowed operators.
It is common practice to extrapolate the matched results to $t \rightarrow 0$.
The extrapolation is usually very smooth and it removes any residual flow time dependence.
Only a thorough numerical investigation can confirm these expectations, but it is 
conceivable that the use of the ratios $R_n^h(t)$ will be beneficial also 
to weaken the contributions from higher dimensional operators.
In conclusion, choosing a renormalization scale of $\mu=2$ GeV, 
it is possible to determine ratios of matrix elements 
where the perturbative corrections are very small,
and where the higher order contributions in flow time 
could be smoothly extrapolated to zero.

\end{document}